\documentstyle [12pt,a4p,epsfig,amsmath,multicol]{article}
\begin{document}
\vspace*{0.6cm}

\begin{center} 
{\normalsize\bf Ruler measurements give space-time-transformation-independent
  invariant lengths}
\end{center}
\vspace*{0.6cm}
\centerline{\footnotesize J.H.Field}
\baselineskip=13pt
\centerline{\footnotesize\it D\'{e}partement de Physique Nucl\'{e}aire et 
 Corpusculaire, Universit\'{e} de Gen\`{e}ve}
\baselineskip=12pt
\centerline{\footnotesize\it 24, quai Ernest-Ansermet CH-1211Gen\`{e}ve 4. }
\centerline{\footnotesize E-mail: john.field@cern.ch}
\baselineskip=13pt
\vspace*{0.9cm}
\abstract{Two thought experiments are described in which ruler measurements of spatial 
 intervals are performed in different reference frames. They demonstrate that such intervals
 are frame-independent as well as independent of the nature of the space-time transformation
 equations. As explained in detail elsewhere, the `length contraction' effect of
 conventional special relativity theory is therefore spurious and unphysical.}
\vspace*{0.9cm}
\normalsize\baselineskip=15pt
\setcounter{footnote}{0}
\renewcommand{\thefootnote}{\alph{footnote}}

    The concept of a `ruler measurement' of a spatial interval is a very familiar one in the everyday world.
    A `ruler' is a flat object with a rectilinear boundary furnished with equally spaced `marks', {\rm M}, specifying
  positions along the boundary, labelled by ordinal numbers: ${\rm M}(I),~I = 1,2,3,...~$. In order to perform
    a length measurement, `pointers' on the objects, the spatial separation of which is to be determined,
     are placed against the ruler and the closest marks to the pointers ${\rm M}(I),~{\rm M}(J)~(J>I)$ are noted. The
     ruler measurement of the spatial separation of the pointers is then $J-I$ in units of the inter-mark
     separation, with a maximum uncertainty of one unit. A mathematical calculus to rigorously specify such
      measuremenrs, in terms of pointer-mark coincidences (PMC), was proposed in Ref.~\cite{JHFSTP1}, but for
     the purposes of the present paper the simple definition just given is sufficient. 
     \par In the case that the to-be-measured objects are at rest relative to the ruler, time plays no
       role in the measurement. However if they are in uniform or accelerated motion, with the same or different
         velocities parallel to the edge of the ruler, the times at which the PMCs, constituting
       the raw measurements, are observed are important. Two distinct situations are possible:
             \begin{itemize}
            \item[(1)] The objects have different velocities.
             \item[(2)]  The objects have the same (possibly time-varying) velocity at all times.
     \end{itemize}
        In the first case, the spatial interval between the objects changes with time
      and a measurement of the spatial separation at any epoch~\footnote{The word `epoch' denotes a
        particular instant of time in some reference frame. It can be operationally
         defined as a particular PMC corresponding to a spatial coincidence between the moving hand of
        an analogue clock at rest in the frame (the pointer) and a mark on its dial.} requires the {\it simultaneous}
        recording of two PMCs. In case (2), which is that of the examples to be discussed in
        this paper, simultaneous observation of PMCs is also required to define the
         spatial separation of the objects, but the latter is a time-independent quantity.
          Note that the spatial separation in case (2) is a constant ---independent of both
         time and the velocity of the objects--- provided that the latter is the same at all
         epochs. This statement is valid for both uniform and accelerated motion of
         the two objects.
           \par Using the above definitions it will now be shown, using two simple examples,
              that ruler measurements of the spatial separation of two objects undergoing
              similar motion give the same result in both an inertial frame from which
               their motion is observed, or in the co-moving reference frame of the
               two objects. Furthermore, this equality is independent of the transformation
               equations relating space and time coordinates in the two frames. A corollary
             is that the `length contraction' and `relativity of simultaneity' effects of
            conventional special relativity are illusory~\cite{JHFFJMP1,JHFCRCS,JHFACOORD,JHFUMC}.
      \par The first example is the thought experiment shown in Fig.~1. Two `pointer trams',
            PT1 and PT2, are at rest on straight tracks aligned with two mark poles M3 and M4 
           respectively, separated by the distance $L$. The ruler measurement of the
        initial separation of the trams is thus $L$. Two further mark poles, M1 and M2, are displaced
         from M3 and M4 respectively by a distance $3L$ in the direction of motion
          of the trams. All of the mark poles are equipped with lamps. Mark poles M3 and M4 are 
          also equipped with local synchronised clocks connected to the lamps and the system controlling
          the motion of the adjacent tram. Because the clocks are at rest in the same inertial frame, they
          may be synchronised by any convenient method; for example by `pointer transport' as described in
          Ref.~\cite{JHFSTP1}, or by clock transport from a suitably placed master clock equidistant from
          M3 and M4. On the assumption of light speed isotropy in the frame of the mark poles the clocks
          may also be synchronised by exchange of light signals, according to the Einstein procedure~\cite{Ein1}.
         Even without assuming light speed isotropy, the clocks may be externally synchronised by reception
         of essentially parallel-moving light signals transmitted simultaneously by a distant source
         equidistant from M3 and M4. In this case the clocks are stopped with the same setting, and are
         started on receipt of the signals. Since the clocks and the signal source are at rest in the
         same frame during the synchronisation procedure there is no necessity to consider any
         putative `relativity of simultaneity' which may occur when synchronised clocks are 
         observed in different inertial frames.
          \par The lamps attached to M3 and M4 flash when
         the trains start to move, in an identical manner, down the tracks while
         M1 and M2 flash at the epoch when the front ends of the moving trams are aligned 
        with them. The simultaneity of the signals of the lamps on M1 and M2 or M3 and M4 endorses
        the validity of the corresponding ruler measurements.
\begin{figure}[htbp]
\begin{center}\hspace*{-0.5cm}\mbox{
\epsfysize12.0cm\epsffile{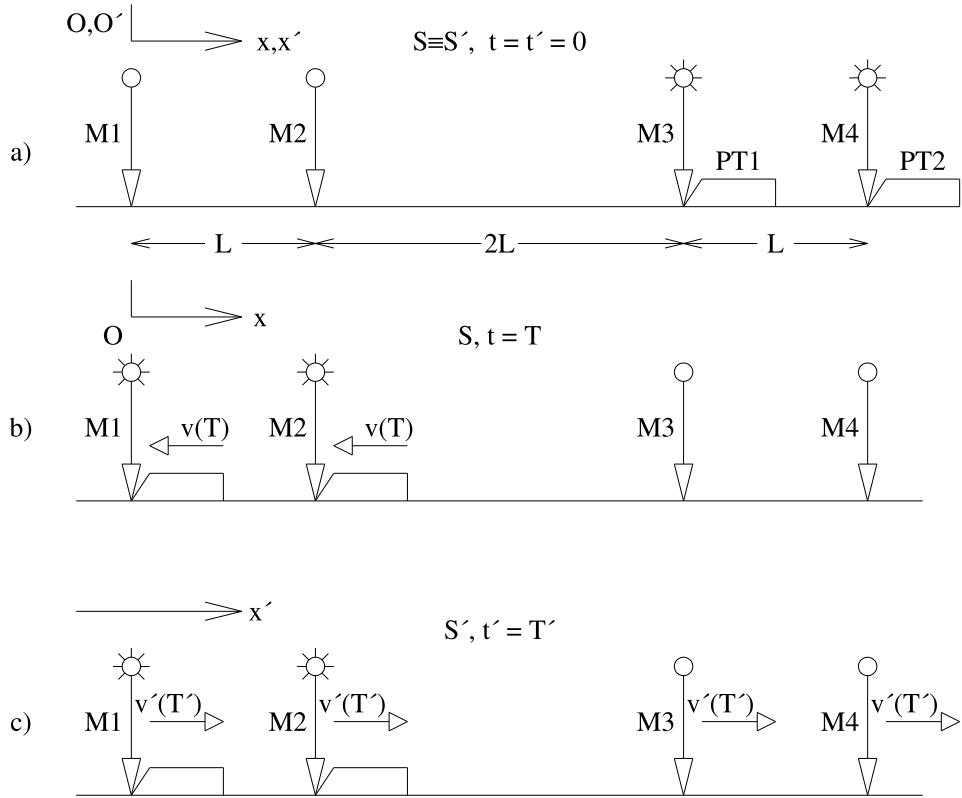}}
\caption{{\em Spatial configurations in frames S [a),b)] and S'[a),c)] of a thought experiment
     in which the pointer trams PT1 and PT2 are simultaneously accelerated in a similar manner so as to pass
    from positions of spatial coincidence with stationary mark poles M3 and M4 at $t = t' = 0$ [a)] 
    to simultaneous spatial coincidence with M1 and M2 at time T in S [b)] or time T' in S' [c)].
    Time-coincident events are signalled by simultaneous flashes of lamps on M3 and M4 in a) and
    on M1 and M2 in b) and c). See text for discussion.}}
\label{fig-fig1}
\end{center}
\end{figure}
          \par Configurations observed in 
          the rest frame, S, of the mark poles are shown in Figs.~1a and 1b. At $t = 0$, as recorded
           by a clock at rest in S, synchronous signals from the local clocks at  M3 and M4 cause the lamps at M3 and M4
           to flash and PT1 and PT2 to start to move towards
           M1 and M2. At time $t = T$, when the instantaneous velocity of PT1 and PT2 is
           $v(T)$, the front of PT1 is aligned with M1 and the  front of PT2 is aligned with M2.
            At this instant the lamps on M1 and M2 flash. The measured spatial separation of the
            two trams is then $L$ ---the same as when they were at rest--- demonstrating the time-independence
            of their separation for case (2).
          \par The same journey of the trams between the mark poles, as observed in the co-moving frame S'
          of the trams, is shown in Figs.~1a and 1c. If $t'$ is the epoch recorded by a clock at rest in
        the frame S', then at $t = t' = 0$, S' is identical to S, so the configuration is again that
         of Fig.~1a. Observers at rest in the trams will see the mark poles M1 and M2 start to move
            simultaneously towards them, with identical motion, so that at the time $t' = T'$, when their
            observed velocity is $v'(T')$, they arrive simultaneously at the front ends of PT1 and PT2
             respectively when their lamps flash (Fig~1c). For this ruler measurement, showing that
              the separation of M1 and M2, as measured in S', is also $L$, the roles of the trams
              and mark poles are inverted. The ends of the trams constitute the marks
               and the moving poles the pointers. It is evident that the flashes of the lamps
              on M1 and M2 are simultaneous in both S and S' ---there is no `relativity of
              simultaneity' effect
             \par The conclusions of the thought experiment ---the invariance of the
               results of ruler measurements of the spatial separations of PT1 and PT2
                 or M1 and M2 performed in different frames in relative motion and the
                 absence of any relativity of simultaneity effect--- follow only from 
                  the initial postulate of identical motion of the trams in the frame S, and
                  are independent of the form of the space-time transformation equations
                    relating $t'$ to $t$. Suppose the acceleration of PT1 or PT2 is some
                 arbitary function of $t$: $a(t)$. The velocity at epoch $t$ of PT1 or PT2
           is then
        \begin{equation}
           v(t) = \int_0^t a(t') dt' 
        \end{equation}
         and the spatial displacement, $d(t)$, at epoch $t$ is
          \begin{equation}
           d(t) = \int_0^t v(t'') dt'' =  \int_0^t dt'' \int_0^{t''} a(t') dt'  
        \end{equation}
           The spatial separations $\Delta x_{PT}(t)$ and $\Delta x'_{PT}(t')$ in S and S' 
          respectively are, from the geometry of Fig.~1:
          \begin{eqnarray}
          \Delta x_{PT}(t) & \equiv & x({\rm PT2})-x({\rm PT1}) \nonumber \\
                           & = & [4L-d(t)]-[3L-d(t)] \nonumber \\   
                           & = & L = \Delta x_{PT}(0) =  \Delta x'_{PT}(t')
               \end{eqnarray}
     Space-time transformation equations yield from $a(t)$ the corresponding acceleration $a'(t')$
      of M1 and M2 as observed in the frame S'. Similarly to (1) and (2) above it follows that
          \begin{equation}
           v'(t') = \int_0^{t'} a'(t''') dt''' 
        \end{equation}
          \begin{equation}
           d'(t') = \int_0^{t'} v'(t'') dt'' =  \int_0^{t'} dt'' \int_0^{t''} a'(t''') dt'''  
        \end{equation}      
      The spatial separations  $\Delta x'_{M}(t)$ and $\Delta x_{M}(t)$ of M1 and M2 in S' and S are:
          \begin{eqnarray}
          \Delta x'_{M}(t') & \equiv & x'({\rm M2})-x'({\rm M1}) \nonumber \\
                           & = & [L+d'(t')]-d'(t') \nonumber \\   
                           & = & L = \Delta x'_{M}(0) =  \Delta x_{M}(t)
      \end{eqnarray}
\begin{figure}[htbp]
\begin{center}\hspace*{-0.5cm}\mbox{
\epsfysize18.0cm\epsffile{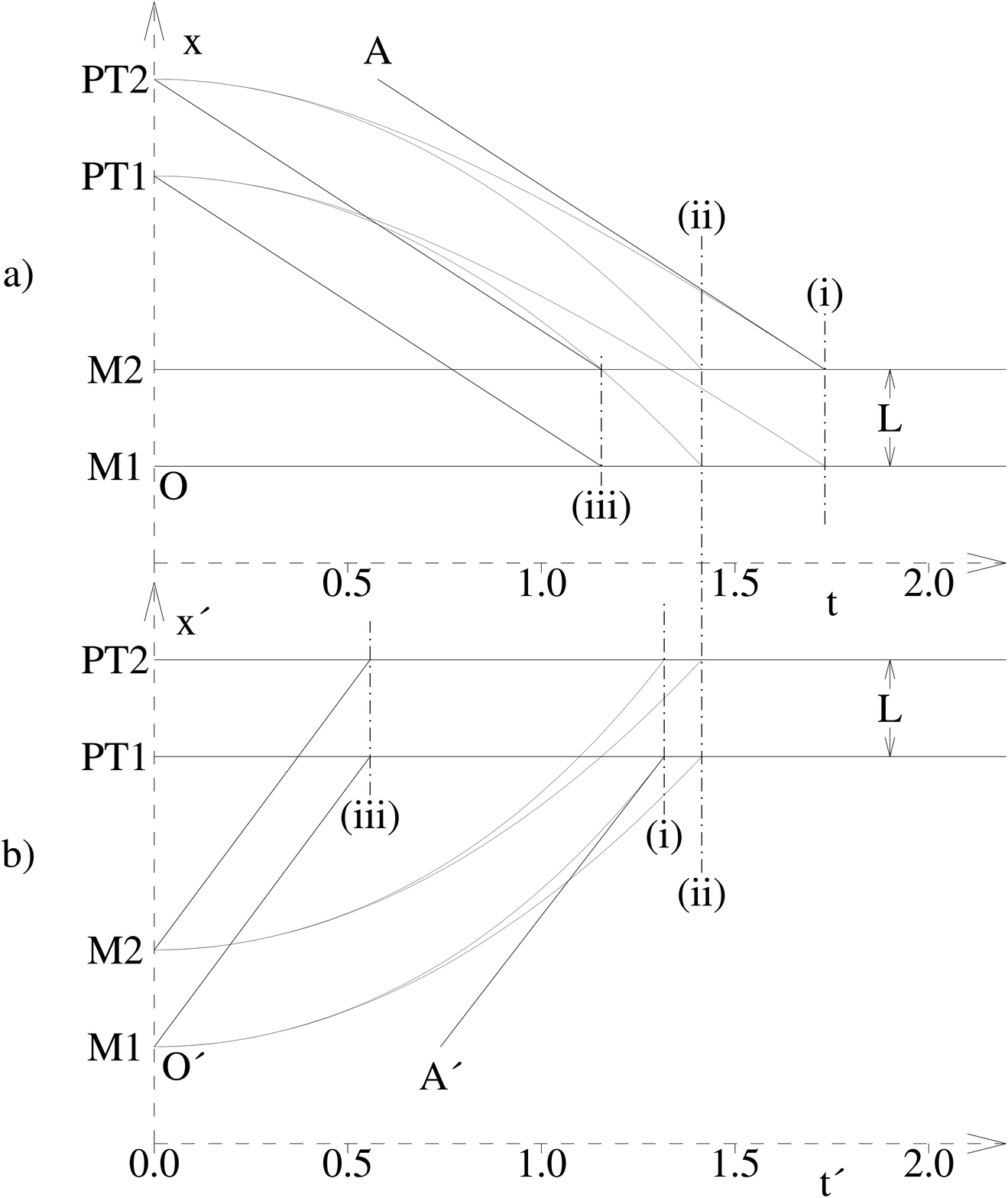}}
\caption{{\em World lines of PT1, PT2, M1 and M2 in S, a) or S', b) for different acceleration
    programs and space-time transformation equations. $c = a_0 = 1$, $L = 1/3$. Ruler measurements
   of the spatial separations of  PT1 and PT2 or M1 and M2 are indicated by the vertical
   dot-dashed lines for the three different cases considered. See text for discussion.}}
\label{fig-fig2}
\end{center}
\end{figure}
       \par The world lines of PT1,PT2,M1 and M2 in S and S' for different acceleration programs and space-time
         transformation equations are shown in Fig.~2 for the following examples:
     \begin{itemize}
      \item[(i)] `Parabolic motion' in special relativity~\cite{Born,Sommerfeld,Rindler,Marder,JHFABSIM}:
           \begin{eqnarray} 
            x_{PT1}(t)& = & x_{PT2}(t)-L = 3L-\frac{c^2}{a_0}\left[\sqrt{1
             +\left(\frac{a_0 t}{c}\right)^2}-1\right] \\
     x'_{M1}(t')& = & x'_{M2}(t')-L = \frac{c^2}{a_0}\left[{\rm cosh}\frac{a_0 t'}{c}-1\right] 
       \end{eqnarray}
       \item[(ii)] Constant acceleration in Galilean relativity:
          \begin{eqnarray} 
            x_{PT1}(t)& = & x_{PT2}(t)-L = 3L-\frac{1}{2}a_0 t^2 \\
     x'_{M1}(t)& = & x'_{M2}(t)-L = \frac{1}{2}a_0 t'^2
         \end{eqnarray}
           These world lines are the $c \rightarrow \infty$ limits of (7) and (8)
      \item[(iii)] Uniform motion after impulsive acceleration in special relativity:
                \begin{eqnarray} 
            x_{PT1}(t)& = & x_{PT2}(t)-L = 3L-vt \\
     x'_{M1}(t)& = & x'_{M2}(t)-L = v't'= \gamma v t
          \end{eqnarray} 
         where $\gamma \equiv \sqrt{1-(v/c)^2}$. The constant velocity, $v$, is chosen according to
      the equations~\cite{JHFABSIM}:
            \begin{eqnarray} 
             T & = & \frac{c}{a_0}\sqrt{\left(1 +\frac{3 a_0 L}{c}\right)-1} \\
             v &  = & -\frac{a_0 T}{\sqrt{1 +\left(\frac{a_0 T}{c}\right)^2}} = -\frac{a_0 T}{\gamma(T)} \\
             v'& = & c{\rm sinh}\frac{a_0 T'}{c} = a_0 T = -\gamma(T)v(T)
    \end{eqnarray} 
              In this way, PT1 and PT2 have the same velocity after impulsive acceleration as they have
              when arriving at M1 and M2 in case (i) above. It corresponds to the limits
              $a_0 \rightarrow \infty$, $T  \rightarrow 0$ in Eqs.~(14) and (15) for finite values of $a_0 T$,
              $v$ and $v'$.
         \end{itemize}
       Since the $c \rightarrow \infty$ (Galilean) and $t,t'\rightarrow 0$ limit of (7) and (8) are
       the same, for the choice of units and parameters $c = a_0 = 1$, $L = 1/3$ of Fig.~2, the world
       lines of PT1,PT2 in S (Fig.~2a) and of M1,M2 in S'(Fig.~2b) are indistinguishable, for $t,t' < 0.5$
       between cases (i) and (ii). For case (ii) the shapes of the world lines of M1 and M2 in S' are
        mirror images of those of PT1 or PT2 in S. The different shapes of the world lines of M1 and M2 in
        S' to those of PT1 or PT2 in S for case (i) are due to the time dilation effect. In fact, at corresponding
       values of $t$ and $t'$, the slopes of the world lines of M1 and M2 in S' are $\gamma$ times the slopes of
       of PT1 or PT2 in S~\cite{JHFABSIM}. The straight world lines of PT1 or PT2 in S and of M1 and M2 in S'
         for case (iii)
       have, due to the time dilation effect, slopes in the ratio 1 : $\gamma$, where $\gamma = 2$ corresponding to
       $v =c\sqrt{3}/2$ or  $c = 1$, $a_0 T = \sqrt{3}$ in Eqs.~(14) and (15). 
       \par The lines A (A'), which are tangents to the world lines of PT2 in Fig.~2a and M1 in Fig.~2b, for case (i),
       where they
        intersect those of M2 and PT1 respectively, are parallel to the world lines
        of PT1 or PT2 (M1 or M2) for case (iii).
        \par The spatial separation of PT1 and PT2 or M1 and M2 is equal to $L$ at all times in both S and S'.
         Ruler measurements of this separation for the three different cases in both frames are 
         indicated by the labelled vertical dot-dashed lines. It can be seen in Fig.~2 that the constancy and
           frame invariance  of the spatial separations of PT1 and PT2 or M1 and M2 is a necessary
         geometrical consequence of the identical shapes of their world lines in each frame in all cases.
         This identity of shape is, in turn, a necessary consequence of the identical nature of the
        acceleration programs to which they are subjected. The invariance of the separation
        is also independent of the form of the space-time transformation equations by which the shape
        of the world lines in S' may be derived from those in S. 
     \par The initial configuration of the second thought experiment is shown in Fig.~3a.
     The measuring rod MR, of length $L$, is used to set the separations of the ruler-mark  
     objects A$_1$, B$_1$, A$_2$, B$_2$, A$_3$ and B$_3$ of the ruler R. The separations
    of  A$_1$-B$_1$, B$_1$-A$_2$, A$_2$-B$_2$ and  A$_3$-B$_3$ are set to $L$, and that
    of B$_2$-A$_3$  to $2L$. The front and back ends of MR (as viewed from the left side of
    the ruler in Fig.~3a) are denoted by F and B respectively. As discussed above,
    if the ends of the moving object are simultaneously 
    aligned with any two of the mark-objects in the proper frame of the ruler,
    the length of the moving object is defined to be equal to the separation of
    the mark-objects in the proper frame of the ruler.
    \par The ruler R is subjected to a similar acceleration program to that in case (i) above,
     specified by the parameter $a_0$, in 
     such a sense that, in the proper frame, S' of R, MR is observed to move to the right.
      The equations describing the world lines of the ends of MR as observed in S', are then 
      similar to (8) above:
      \begin{equation}
      x'({\rm F})  = x'({\rm B})-L = \frac{c^2}{a_0}\left[\cosh\frac{a_0 t'}{c}-1\right] \\  
      \end{equation}
\begin{figure}[htbp]
\begin{center}\hspace*{-0.5cm}\mbox{
\epsfysize12.0cm\epsffile{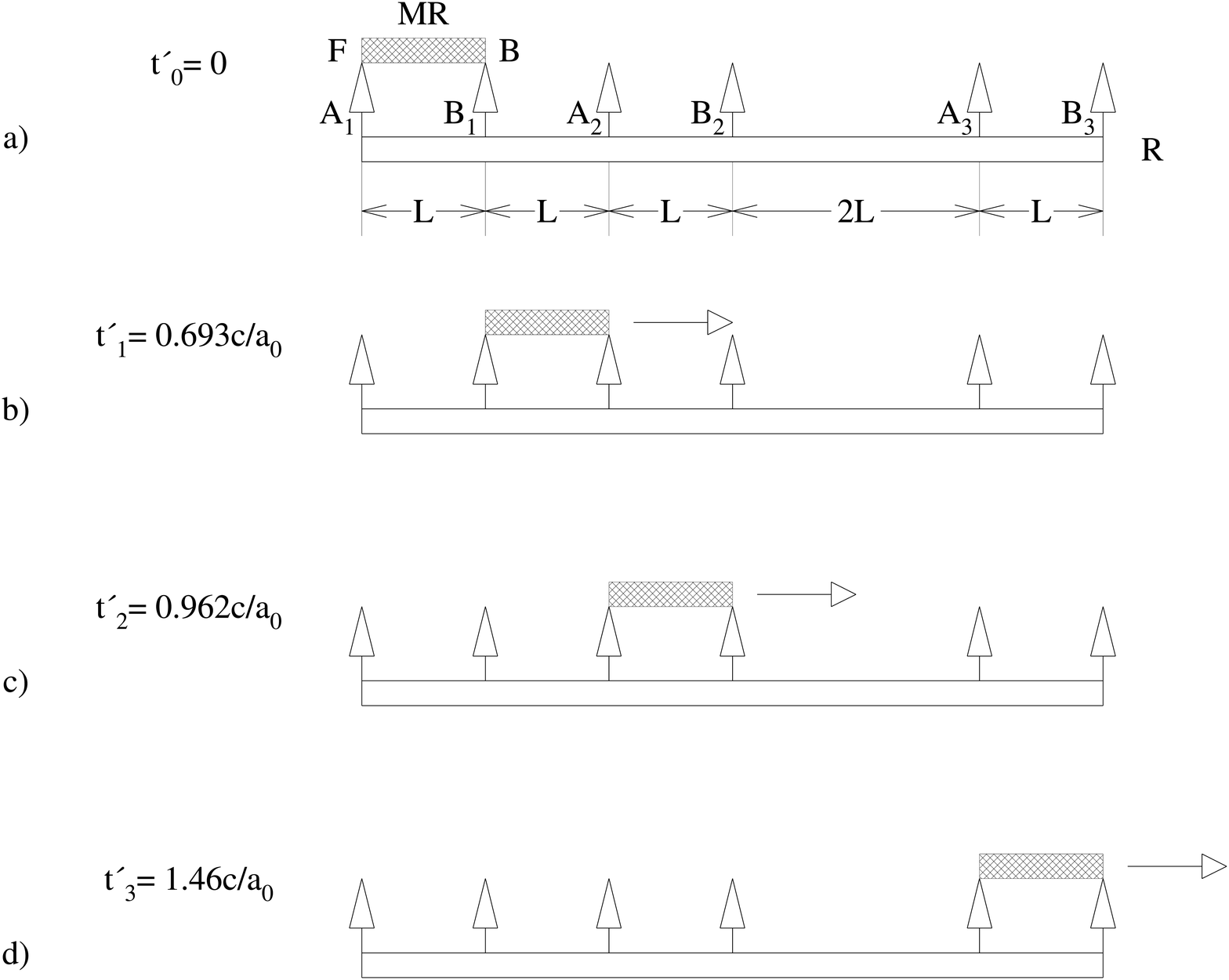}}
\caption{{\em Four measurements of a stationary measuring rod MR are performed by a moving ruler
  R. The measurements as observed in the proper 
  frame, S', of R are shown. In a), MR is at rest relative to R. In b) and c) measurements are
    made while R is accelerating. In d) the ruler moves uniformly relative to MR. In all
    cases MR is measured to have the same length $L$ ---there is no `length contraction' 
     effect.}}
\label{fig-fig3}
\end{center}
\end{figure}

\begin{figure}[htbp]
\begin{center}\hspace*{-0.5cm}\mbox{
\epsfysize15.0cm\epsffile{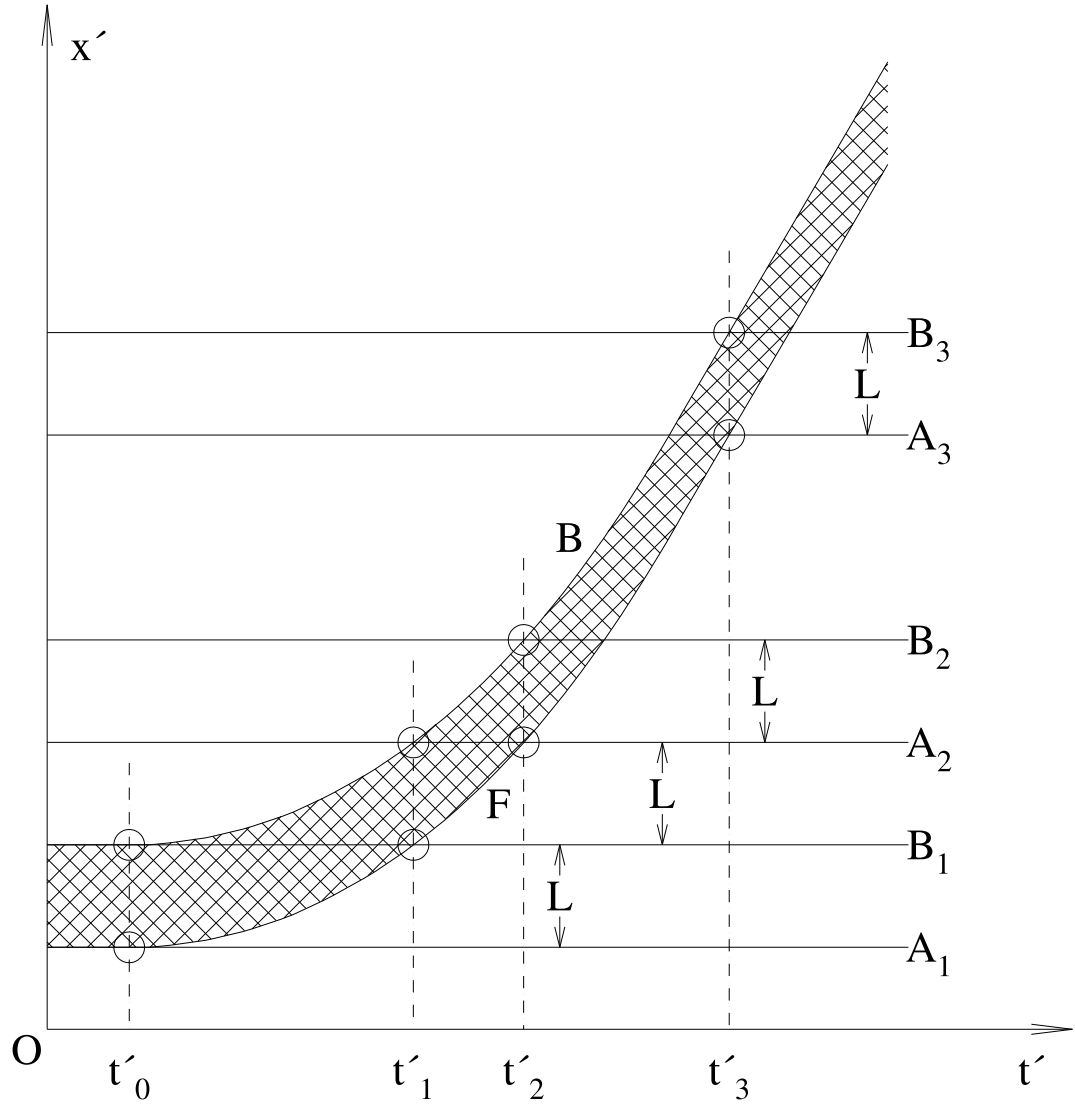}}
\caption{{\em World lines of the front (F) and back (B) ends of the stationary measuring 
  rod MR, as viewed from the proper frame of the ruler R, showing the epochs $t'_0$, $t'_1$,
  $t'_2$ and $t'_3$ of the four concordant measurements of the length of MR as indicated by the vertical
   dot-dashed lines.}}
\label{fig-fig4}
\end{center}
\end{figure}

\begin{figure}[htbp]
\begin{center}\hspace*{-0.5cm}\mbox{
\epsfysize15.0cm\epsffile{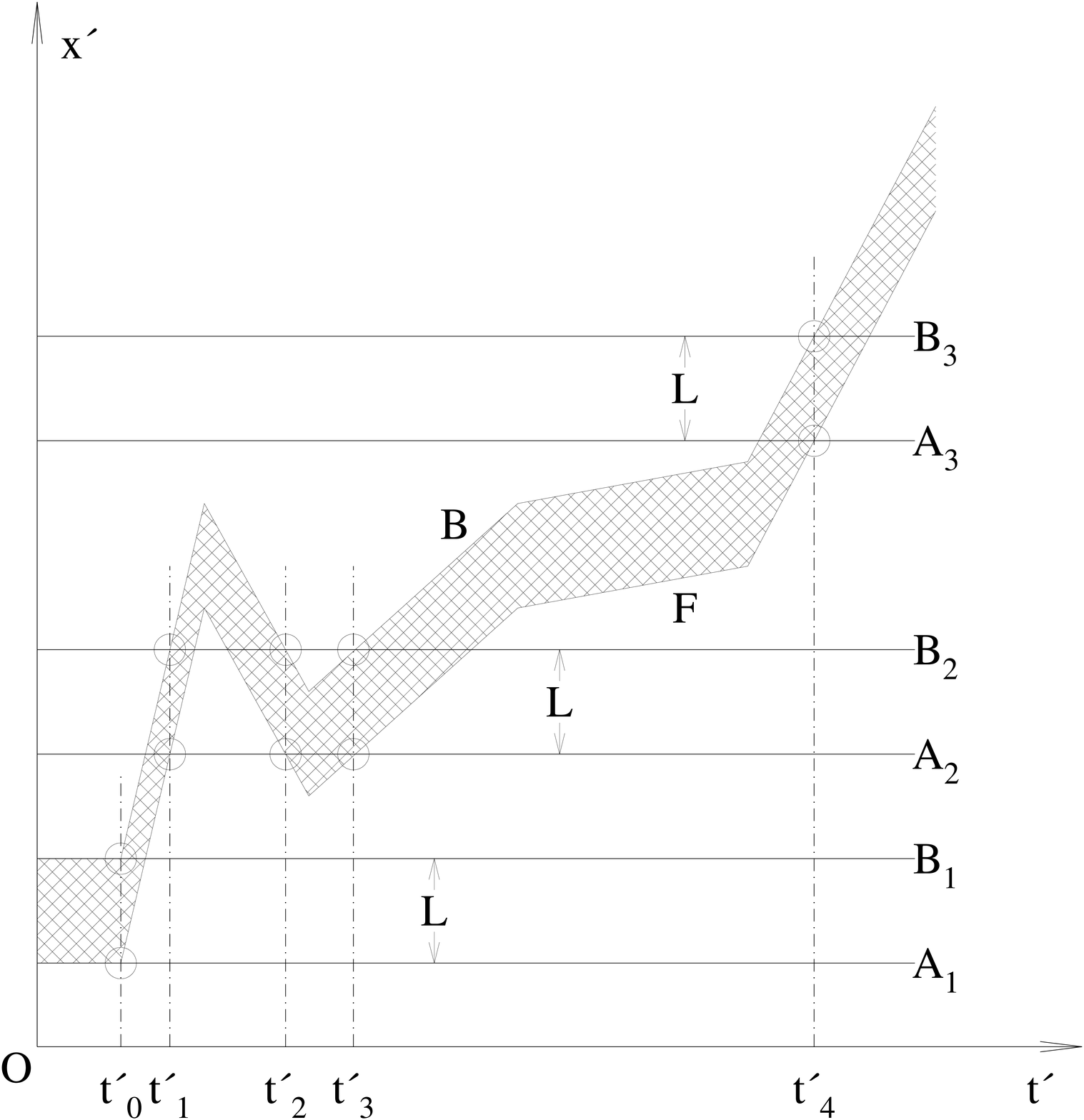}}
\caption{{\em World lines of the front (F) and back (B) ends of the stationary measuring 
  rod MR, as viewed from the proper frame of the ruler R, when the latter is subjected
   to a series of impulsive accelerations. Five concordant measurements of the length 
   of MR, indicated by the vertical
   dot-dashed lines, at epochs $t'_0$, $t'_1$,
  $t'_2$, $t'_3$ and $t'_4$ are obtained}}
\label{fig-fig5}
\end{center}
\end{figure}

       \par In Fig.~3a, the length of MR at rest, $L$, is measured by the separation
       of A$_1$ and B$_1$. It follows from (16) that the length of
       MR is also measured to be $L$ by simultaneous
      coincidence of F and R with B$_1$ and  A$_2$ at the epoch $t'_1 =0.693c/a_0$.
 In the Figs.~3-5, units and dimensions are chosen so that 
     $a_0 = c = 4L = 1$. Similar measurements are performed by  A$_2$ and  B$_2$ at
      $t'_2 =0.962c/a_0$ and by A$_3$ and  B$_3$ at
      $t'_3 =1.46c/a_0$. As in Ref.~\cite{JHFABSIM}, it is assumed that the
     acceleration halts when $x'({\rm F}) = c^2/a_0$ at the epoch in S':
      \begin{equation}
       t'_{acc} = \frac{c}{a_0}{\rm arccosh}(2) = 1.317\frac{c}{a_0}
  \end{equation}
       so that for epochs $t' \ge  t'_{acc}$ , S' is an inertial frame and $v'$ is constant.
     \par The world lines of F and B in S'and the four measurements of the
      length of MR ---one at rest, two during accelerated motion and one during 
   uniform motion--- are plotted in Fig. 4. Again, each concordant ruler measurement is indicated by
    a vertical dot-dashed line. This figure shows clearly that the 
    physical basis for the equality of all the length measurements is that that the first member of (16)
    may be transposed to give:
    \begin{equation} 
  x'({\rm B}, t')- x'({\rm F}, t')  = L
      \end{equation}
  where the $t'$ dependence of each term is shown explicitly. 
   Thus the world line of B is derived from that of F by displacing it by the
   distance $L$ along the positive $x'$-axis. The operation $x' \rightarrow x'+L$
   corresponds to moving the origin of coordinates by a distance $L$. The shape
   of the world line of an end of MR is invariant under this transformation, and
   so manifests translational invariance.
   \par For definiteness, the case of `hyperbolic motion' of the ruler, according to Eqn(16),
    was considered above, but it is clear that the invariance of length measurements
   of MR must be independent of the acceleration program of R. In Fig.~5, for example,
    a series of impulsive accelerations of sign $+,-,+,+$ are applied, yielding 
    equal length measurements at the epochs $t'_0$, $t'_1$, $t'_2$, $t'_3$ and $t'_4$
    as shown. One measurement corresponds to a simultaneous coincidences of F-A$_1$ and B-B$_1$ three
    to simultaneous coincidences F-A$_2$ and B-B$_2$,
     and one to a simultaneous coincidence of  F-A$_3$ and B-B$_3$.
   \par Discussion of the reason for the spurious nature of the correlated `length contraction'
     and `relativity of simultaneity' effects of conventional special
     relativity~\cite{Ein1}
     may be found
     in Refs.~\cite{JHFSTP1,JHFFJMP1,JHFCRCS,JHFACOORD,JHFUMC}. In particular it is shown in
        Ref.~\cite{JHFFJMP1}, recently published in this journal, that these effects
         result from the neglect of certain additive constants in the space-time
         Lorentz transformation equations, mentioned by Einstein~\cite{Ein1}, but not implemented by him,
         that are required to correctly describe synchronised clocks at different spatial
         positions.
   \newline
   {\bf Acknowledgments}
   \par I thank an anonymous referee for pointing out the importance of the concept of synchronisation
         of spatially-separated clocks for correctly specifying the initial conditions of the thought
        experiment shown in Fig.~1.

\end{document}